\documentclass[sigconf, nonacm]{acmart}
\AtBeginDocument{%
  }

\setcopyright{acmlicensed}
\copyrightyear{2018}
\acmYear{2018}
\acmDOI{XXXXXXX.XXXXXXX}
\acmConference[Conference acronym 'XX]{Make sure to enter the correct
  conference title from your rights confirmation email}{June 03--05,
  2018}{Woodstock, NY}
\acmISBN{978-1-4503-XXXX-X/2018/06}

\usepackage{subcaption}
\usepackage{enumitem}

\begin{document}

\title{JARVIS: An Evidence-Grounded Retrieval System for Interpretable Deceptive Reviews Adjudication}

\author{Nan Lu}
\authornote{Both authors contributed equally to this research.}
\authornote{Corresponding author.}
\email{lunan5@jd.com}          
\affiliation{%
  \institution{Beijing Jiaotong University}
  \city{Beijing}
  \country{China}
}
\affiliation{%
  \institution{JD.com}
  \city{Beijing}
  \country{China}
}

\author{Leyang Li}
\authornotemark[1]
\email{lileyang.tokki@jd.com}
\affiliation{%
  \institution{JD.com}
  \city{Beijing}
  \country{China}
}

\author{Yurong Hu}
\email{huyurong7@jd.com}
\affiliation{%
  \institution{JD.com}
  \city{Beijing}
  \country{China}
}

\author{Rui Lin}
\email{linrui@jd.com}
\affiliation{%
  \institution{JD.com}
  \city{Beijing}
  \country{China}
}

\author{Shaoyi Xu}
\authornotemark[2]
\email{shyxu@bjtu.edu.cn}
\affiliation{%
  \institution{Beijing Jiaotong University}
  \city{Beijing}
  \country{China}
}

\renewcommand{\shortauthors}{Nan Lu, Leyang Li, Yurong Hu, Rui Lin, and Shaoyi Xu}
%% No italics, no superscripts, not anonymous
%% Use footnote or author note to identify equal contribution and/or contact author info

\begin{abstract}
  Deceptive reviews are fabricated feedback designed to artificially manipulate the perceived quality of products. Within modern e-commerce ecosystems, these reviews remain a critical governance challenge. Despite advances in review-level and graph-level detection methods, two pivotal limitations remain: inadequate generalization and lack of interpretability. To address these challenges, we propose \textbf{JARVIS}, a framework providing \textbf{J}udgment via \textbf{A}ugmented \textbf{R}etrieval and e\textbf{VI}dence graph \textbf{S}tructures. Starting from the review to be evaluated, it retrieves semantically similar evidence via hybrid dense–sparse multimodal retrieval, expands relational signals through shared entities, and constructs a heterogeneous evidence graph. The large language model then performs evidence-grounded adjudication to produce interpretable risk assessments. Offline experiments demonstrate that JARVIS enhances performance on our constructed review dataset, achieving a precision increase from 0.953 to 0.988 and a recall boost from 0.830 to 0.901. In the production environment, our framework achieves a 27\% increase in the recall volume and reduces manual inspection time by 75\%. Furthermore, the adoption rate of the model-generated analysis reaches 96.4\%.
\end{abstract}

\begin{CCSXML}
<ccs2012>
<concept>
<concept_id>10002951.10003317.10003371.10003386</concept_id>
<concept_desc>Information systems~Multimedia and multimodal retrieval</concept_desc>
<concept_significance>500</concept_significance>
</concept>
<concept>
<concept_id>10002951.10003317.10003347</concept_id>
<concept_desc>Information systems~Retrieval tasks and goals</concept_desc>
<concept_significance>500</concept_significance>
</concept>
<concept>
<concept_id>10002951.10003317.10003338.10003341</concept_id>
<concept_desc>Information systems~Language models</concept_desc>
<concept_significance>500</concept_significance>
</concept>
</ccs2012>
\end{CCSXML}

\ccsdesc[500]{Information systems~Retrieval tasks and goals}
\ccsdesc[500]{Information systems~Multimedia and multimodal retrieval}
\ccsdesc[500]{Information systems~Language models}

\keywords{Fraud Detection, Multi-Modal Retrieval, Heterogeneous Graph, Large Language Model}


\maketitle

\section{Introduction}
Online reviews play a significant role in shaping consumer purchasing decisions and influencing product reputation on e-commerce platforms\cite{tadelis2016reputation}. To enhance the influence and reputation of their products, many merchants resort to manipulating online reviews\cite{gossling2018manager, lee2018sentiment}. Such manipulative practices create a terrible scenario: in the short term, users are deceived into purchasing sub-optimal products, leading to financial and emotional dissatisfaction. From a long-term perspective, deceptive reviews erode the consumer trust, ultimately undermining the platform's ecosystem and viability.

Despite years of research dedicated to deceptive review detection, existing methods suffer from two major limitations (Section \ref{sec_2}): they either lack generalization by focusing solely on individual review features, or they fail to provide a clear reasoning chain for their decisions, necessitating heavy reliance on manual re-verification.

\begin{figure*}[htbp]
  \includegraphics[width=\textwidth]{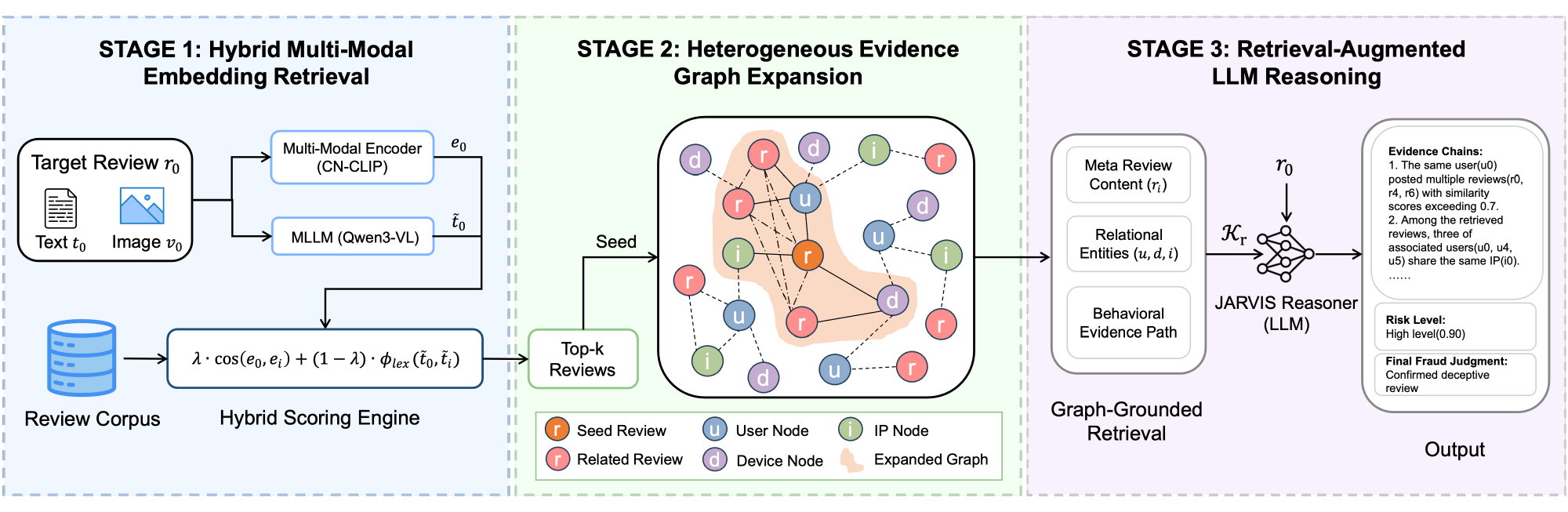}
  \caption{The overall architecture of the proposed deceptive review detection framework. The process begins with Stage 1, encoding the target review via dual multi-modal encoders and a hybrid scoring engine for retrieval. In Stage 2, the retrieved Top-k reviews seed a heterogeneous graph expansion, where the dash-dot line denotes $\mathcal{E}_{rr}$, the dashed line represents $\mathcal{E}_{re}$, and the solid line indicates $\mathcal{E}_{ee}$. Finally, Stage 3 feeds review content, relational entities, and behavioral paths into the LLM based reasoner for graph-grounded evidence chain generation and fraud classification.}
  \Description{Enjoying the baseball game from the third-base
  seats. Ichiro Suzuki preparing to bat.}
  \label{fig:teaser}
\end{figure*}

To address these limitations, we propose a framework named \textbf{JARVIS}, providing \textbf{J}udgment via \textbf{A}ugmented \textbf{R}etrieval and e\textbf{VI}dence graph \textbf{S}tructures. Our method addresses the issue of poor generalization by employing a hybrid multi-modal retrieval strategy to obtain reviews that are structurally or semantically similar to the target. Furthermore, by constructing a heterogeneous evidence subgraph, we provide the Large Language Model with the essential contextual sources required for reasoning, thereby effectively resolving the lack of an interpretable evidence chain. The entire pipeline is training-free: it takes review images and text as input and outputs a definitive adjudication accompanied by a complete evidence chain. Both our offline and online experimental results demonstrate that our approach significantly outperforms existing baselines. Our contributions are summarized as follows:
\begin{itemize}[nosep, leftmargin=*]
    \item We systematically analyze and summarize existing deceptive review detection methods, identifying two major critical limitations. Based on these insights, we propose a retrieval-driven and interpretable framework for review fraud detection.
    \item We propose a retrieval methodology that decomposes multi-modal reviews into dense and sparse representations, which are then organically integrated to perform hybrid retrieval.
    \item Our approach significantly outperforms the baseline in both offline benchmarks and online A/B tests, delivering tangible value to production environments.
\end{itemize}
\vspace{-6pt}

\section{Literature Review}
\label{sec_2}
\noindent\textbf{Deceptive Reviews Detection.} 
Existing research on deceptive review detection predominantly follows two paradigms: review-level and graph-level. Review-level methods focus on discriminating fraudulent content by integrating textual styles, sentiment cues, and multi-modal information using both classical machine learning\cite{alsubari2022data, tufail2022effect, abdulqader2022fake, khurshid2018enactment, jalther2019reputation, khatun2025leveraging} and deep neural networks\cite{jing2022semi, csenol2025domain, liu2022detection, crawford2021using, mohawesh2024fake, veluru2025multimodal, liu2019opinion}. While recent studies\cite{liu2025detecting, gambetti2023combat} have leveraged Large Language Models (LLMs) for their superior semantic reasoning, these efforts primarily target "AI-generated reviews", thus overlooking broader manipulative patterns and remaining inherently review-level. In contrast, graph-level approaches \cite{manaskasemsak2023fake, ren2022research, he2022detecting, wang2025saft, zhang2025dual, li2019spam} treat the problem as a structural anomaly detection task, modeling the complex relational web among users, products, and timestamps to uncover large-scale collusive behaviors.

Despite their respective strengths, both paradigms suffer from limitations in real-world deployment. Review-level models are easily circumvented by sophisticated rewriting and lack the context to capture fraudulent signals beyond the review. Conversely, while graph-level models excel at detecting group anomalies, they often function as "black boxes", struggling to provide the fine-grained, interpretable evidence. These limitations highlight the need for a unified framework that can jointly support review-centric analysis, scalable evidence retrieval, and interpretable decision-making, which motivates our work.

\noindent\textbf{Multi-Modal Embedding Retrieval} has been successfully deployed across various domains, including search engines\cite{liu2025multimodal}, recommendation systems\cite{wang2025mmsrarec}, and risk management\cite{liang2025embedding}. In this work, we adopt a hybrid retrieval approach that combines dense and sparse representations allowing the framework to simultaneously address semantic paraphrasing and the presence of rare characters.

% \noindent\textbf{LLMs in Risk Management and Control.} Recent advances in Large Language Models (LLMs) have led to their widespread adoption across various risk management and control scenarios, driven by their strong capabilities in semantic understanding, contextual reasoning, and flexible knowledge integration. Yang et al. established a multi-round evaluation benchmark named Fraud-R1\cite{yang2025fraud} to assess the proficiency of LLMs in anti-fraud tasks and user safety protection. Lu et al. proposed the SHERLOCK\cite{lu2025sherlock} framework to assist risk analysts in the adjudication and investigation of high-risk transaction orders. Tan et al.\cite{tan2025understanding}, on the other hand, focused on the application of LLMs for the adjudication and forensic analysis of financial fraud cases. However, to the best of our knowledge, the application of LLMs to deceptive review detection remains largely underexplored, let alone their integration with structured evidence for interpretable risk analysis. Therefore, our work seeks to leverage LLMs in conjunction with an external review-based Knowledge Graph (KG) to identify residual risks and facilitate expert adjudication within the domain of transactional risk control.

\section{Method}
We propose JARVIS, a retrieval-driven framework for interpretable deceptive review adjudication. The framework consists of three core components: (1) Hybrid Multi-Modal Embedding Retrieval, (2) Heterogeneous Evidence Graph Expansion, and (3) Retrieval-Augmented LLM Reasoning.
In the following sections, we provide a detailed explanation of these three components.
\subsection{Hybrid Multi-Modal Embedding Retrieval}
This part serves as a critical prerequisite for constructing the heterogeneous evidence graph. Our goal is to retrieve a set of reviews that are semantically similar to the target review, forming an evidence pool that supports subsequent graph expansion. 
% To this end, we initiate retrieval from the meta-review to identify additional evidence exhibiting semantic or structural similarity.
However, this objective introduces several retrieval challenges. First, deceptive reviews often employ domain-specific promotional terminology that is poorly generalized by dense semantic encoders. In parallel, intentional lexical substitutions weaken sparse retrieval signals based on exact term matching. Second, visual content may contain reused templates or promotional patterns that are not fully reflected in textual parts of reviews.

Therefore, we design a \textbf{hybrid multi-modal embedding retrieval module} that integrates dense semantic similarity, sparse lexical matching, and multi-modal description augmentation to achieve high-recall and high-precision evidence discovery.

Let $r_0$ denote the target review, consisting of textual content $t_0$ and an associated image $v_0$. Given a review corpus $\mathcal{R} = \{r_i\}$, our goal is to retrieve a set of semantically and lexically related reviews $\mathcal{C}(r_0) \subset \mathcal{R}$ that can serve as candidate evidence for downstream analysis. To evaluate the similarity scores between pairs of reviews, we employ a hybrid retrieval approach that integrates both dense and sparse retrieval mechanisms:
\begin{equation}
s(r_0, r_i) = \lambda \cdot \cos(\mathbf{e}_0, \mathbf{e}_i) + (1 - \lambda) \cdot \phi_{\text{lex}}(\tilde{t}_0, \tilde{t}_i),
\label{equ:1}
\end{equation}
where $cos(.)$ denotes the cosine similarity between dense multi-modal embeddings, measuring high-level semantic relevance between reviews. $\phi_{\text{lex}}(.)$ represents the sparse lexical matching score computed over augmented textual representations, capturing exact term overlap and salient keyword matching, and $\lambda$ is a hype-parameter controling the trade-off between dense and sparse similarity. $e_i$ is obtained by feeding the textual content of a review and its associated images into a pre-trained multi-modal encoder, which projects these heterogeneous inputs into a unified embedding space. $\tilde{t}_i$ denotes the augmented textual representation derived from $\tilde{t}_i = t_i \cup g_{\text{MLLM}}(v_i)$, where $g_{\text{MLLM}}(.)$ generates a detailed textual description of the image.

By selecting the top-k most similar reviews, we finally get the target set of candidates:
\begin{equation}
\mathcal{C}(r_0) = \text{TopK}_{r_i \in \mathcal{R}} (s(r_0, r_i))
\end{equation}

\subsection{Heterogeneous Evidence Graph Expansion}
Building upon the retrieved candidate reviews, we construct a heterogeneous evidence subgraph to organize cross-node signals. We initialize the seed set as $S_r = {r_0} \cup \mathcal{C}(r_0)$ and similarities between initial nodes as $E_{rr}$. Starting from $S_r$, we expand to entity nodes (e.g., users, devices, IPs) through behavioral relations (e.g., users post the reviews), and collect additional reviews attached to these entities, while retaining review–review similarity links inherited from embedding retrieval. The expanded node set is defined as:
\begin{equation}
\mathcal{V}_r = \mathcal{S}_r \cup \mathcal{N}_E(\mathcal{S}_r) \cup \{ \mathcal{N}_R(\mathcal{N}_E(\mathcal{S}_r)) : |T_r - T_s| \le \Delta T \},
\label{vr}
\end{equation}
where $\mathcal{N}_E$ expands to entities and  $\mathcal{N}_R$ additionally retrieves other reviews associated with these entities. Notably, we impose a constraint on the absolute time difference between reviews belonging to the same entity. This temporal filtering mechanism prevents the introduction of excessive noise from irrelevant reviews. 

The edge set within the subgraph is denoted by $E_r$, which characterizes the multi-relational dependencies between nodes. Specifically, these edges are categorized into three types: (1) $\mathcal{E}_{rr}$, representing the similarities between review nodes. The similarities between the review nodes expanded through entities have not been pre-computed. Therefore, we utilize Equation \ref{equ:1} again to supplement these calculations. (2) $\mathcal{E}_{re}$, representing the associations between entities and review nodes, such as a review being affiliated with a specific product. (3) $\mathcal{E}_{ee}$, representing the connections between entity nodes, such as a user logging into a particular device.
% \begin{equation}
% \mathcal{E}_{rr} \leftarrow \mathcal{E}_{rr} \cup \left\{ s(r_i, r_j) \mid r_i, r_j \in \mathcal{V}_r, \, i \neq j \right\}
% \label{equ4}
% \end{equation}
The final evidence subgraph can be represented as:
\begin{equation}
\mathcal{G}_r = (\mathcal{V}_r, \mathcal{E}_r)
\end{equation}

\subsection{Retrieval-Augmented LLM Reasoning}
We adopt a graph-grounded retrieval strategy, allowing the LLM to query externalized knowledge from $\mathcal{G}_r$ for reasoning. Given a query review $r_0$, the LLM retrieves relevant evidence paths from the graph:
\begin{equation}
\mathcal{K}_r = \text{Retrieve}(r_0, \mathcal{G}_r),
\end{equation}

\begin{table}[htbp]
\centering
\small
\setlength{\tabcolsep}{3pt}
\renewcommand{\arraystretch}{1.1}

\begin{tabular}{p{2.2cm} p{5.4cm}}
\hline
\textbf{Component} & \textbf{Description} \\
\hline

Role Definition &
The LLM is assigned as a senior e-commerce risk control expert and anti-fraud auditor, responsible for analyzing suspicious reviews based on retrieved knowledge graph evidence. \\

\hline

Task Objective &
Given the meta-review and its associated knowledge graph evidence, the model determines whether the review is deceptive. \\

\hline

Graph-Retrieved Evidence &
Evidence is retrieved from the heterogeneous subgraph $\mathcal{G}_r$ \\
&
\quad • Meta-review content (text, image description) \\

&
\quad • Retrieved reviews and acquisition paths \\

&
\quad • Similarities between reviews \\
&
\quad • Entities (user/device/IP/item) \\

&
\quad • Entity-sharing interaction records \\

\hline

Reasoning
&
(1) \textbf{Entity Consistency Audit:} \\
&
Assess whether shared entity activities exhibit abnormal collusion patterns. \\

&
(2) \textbf{Semantic Style Alignment:} \\

&
Examine whether retrieved reviews show unnatural stylistic coordination. \\

\hline

Output Format &
• Evidence chains \\

&
• Risk level (with risk score) \\

&
• Final fraud judgment \\

\hline

\end{tabular}

\caption{Graph-grounded retrieval and reasoning schema for LLM-based review risk assessment. This schema illustrates how JARVIS integrates heterogeneous evidence into a unified LLM reasoning process. By simulating a senior anti-fraud expert, the model transforms raw graph-retrieved data into structured deception judgments and explainable evidence paths.}
\vspace{-1.6em}
\label{tab:llm_reasoning_schema}
\end{table}

\vspace{-4pt}

The detailed LLM configuration, retrieved evidence schema, and reasoning design are presented in Table \ref{tab:llm_reasoning_schema}. Specifically, the reasoning process mimics the cognitive workflow of a human auditor, moving from low-level data extraction (e.g., IP and device logs) to high-level pattern recognition (e.g., unnatural stylistic coordination among reviews). To this end, we have obtained an interpretable verdict and a comprehensive evidence chain for the target review generated by the LLM. High-risk cases trigger automated disposal, while medium-risk evidence chains and scores serve as references for manual adjudication.

\section{Experiments}
\subsection{Offline Evaluation}
\subsubsection{Datasets and Experiments Setup}
We construct a dataset comprising 300,000 reviews spanning 15 active product categories. We select open-source detection methods to be trained and evaluated on our dataset. For a fair comparison with supervised learning baselines, we split the dataset into training and testing subsets. It is worth noting that while our method directly operates on the testing set without any prior training, the supervised baselines\cite{he2022detecting, li2019spam, gambetti2023combat} are trained on the training set to reach their optimal performance. We also compare our framework against the current production model, which is an ensemble system comprising a BERT-based supervised sub-model and a risk-feature-driven graph sub-model.

We employ $\text{CN-CLIP}_\text{ViT-B/16}$\cite{yang2022chinese} to generate dense embeddings, Qwen3-VL-30B-A3B-Instruct\cite{Qwen3-VL} to describe the review images, and BGE-M3\cite{chen2024bge} for sparse vector representations. The hybrid hyperparameter $\lambda$ is set to 0.5 and top-25 most similar reviews are retrieved. For computational efficiency and temporal relevance, the embedding database only maintains reviews within a sliding 30-day time window. In the LLM reasoning phase, we use Qwen3-30B-A3B\cite{yang2025qwen3} for reasoning and risk adjudication.

\subsubsection{Results}
Table \ref{tab:results} presents the experimental results. The results indicate that our method substantially outperforms state-of-the-art research in deceptive review detection. Also, our approach consistently surpasses the production baseline in both precision and recall.

% \begin{table}[htbp]
% \centering
% \footnotesize

% \setlength{\tabcolsep}{5pt}
% \renewcommand{\arraystretch}{1.1} 

% \resizebox{0.9\columnwidth}{!}{   
% \begin{tabular}{lccc}
% \hline
% Method & Precision & Recall & F1-score \\
% \hline
% Baseline & 0.953 & 0.830 & 0.887 \\
% JARVIS      & \textbf{0.988} & \textbf{0.901} & \textbf{0.942} \\
% \hline
% \end{tabular}
% }
% \caption{Performance Comparison Between Baseline and our proposed framework}
% \vspace{-1.6em}
% \label{tab:offline-performance}
% \end{table}

\begin{table}[!htbp]
    \centering
    \caption{Performance Comparison between Different Methods and Our Proposed Framework. P-baseline is denoted as the current production-deployed system. We standardize the backbone models across all evaluated methods: $\text{CN-CLIP}_\text{ViT-B/16}$\cite{yang2022chinese} is employed for all tasks requiring dense embeddings, and Qwen3-30B-A3B\cite{yang2025qwen3} is utilized as the unified LLM.}
    \label{tab:results}
    \begin{tabular}{lccc}
        \toprule
        \textbf{Method} & \textbf{Precision} & \textbf{Recall} & \textbf{F1-score} \\
        \midrule
        Detect\cite{he2022detecting} & 0.803 & 0.815 & 0.809 \\
        Gas\cite{li2019spam}         & 0.917 & 0.692 & 0.743 \\
        Combat\cite{gambetti2023combat} & 0.916 & 0.648 & 0.759 \\
        FraudSquad\cite{liu2025detecting} & 0.686 & 0.601 & 0.641 \\
        \midrule
        P-baseline        & 0.953 & 0.830 & 0.887 \\
        % --- 你的模型 ---
        \textbf{JARVIS} & \textbf{0.988} & \textbf{0.901} & \textbf{0.942} \\
        \bottomrule
    \end{tabular}
\end{table}

\subsection{Ablation Study}
In this section, we conduct ablation studies based on our offline evaluation setup to verify the necessity of each individual module within our proposed framework. Figure \ref{fig:ablation} illustrates the results.

\subsubsection{Retrieval Strategy} We evaluate the performance by considering scenes that exclude dense embeddings, sparse embeddings, and review images, as well as scenarios involving alternative $\lambda$ values. From the results in Figure \ref{fig:ablation-a}, we can conclude that the absence of dense embeddings leads to a significant decline in both precision and recall. In contrast, the removal of image modalities and sparse embeddings primarily results in a notable loss in recall. It is reasonable to hypothesize that reviews with dense semantic similarities play a pivotal role in identifying fraudulent patterns, whereas visual evidence and sparse lexical signals function more as auxiliary cues to expand the coverage of suspicious candidates. Meanwhile, the results also validate that $\lambda = 0.5$ is an optimal setting.

\subsubsection{Graph Node Types} 
\begin{figure}[t]
\centering

\begin{subfigure}[t]{0.49\columnwidth}
    \centering
    \includegraphics[width=1.0\linewidth]{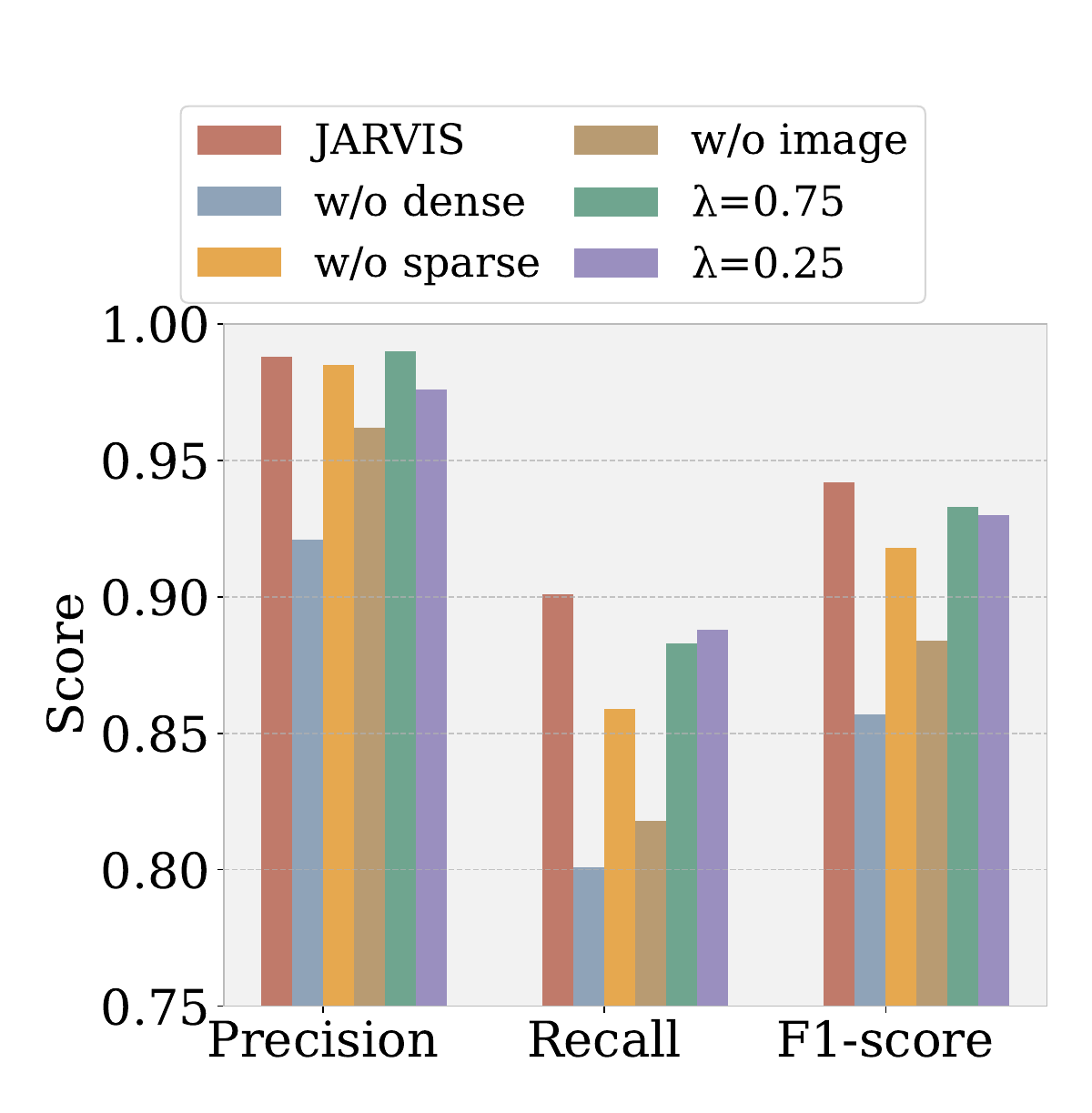}
    \caption{Retrieval Strategy.}
    \label{fig:ablation-a}
\end{subfigure}
\hfill
\begin{subfigure}[t]{0.49\columnwidth}
    \centering
    \includegraphics[width=1.0\linewidth]{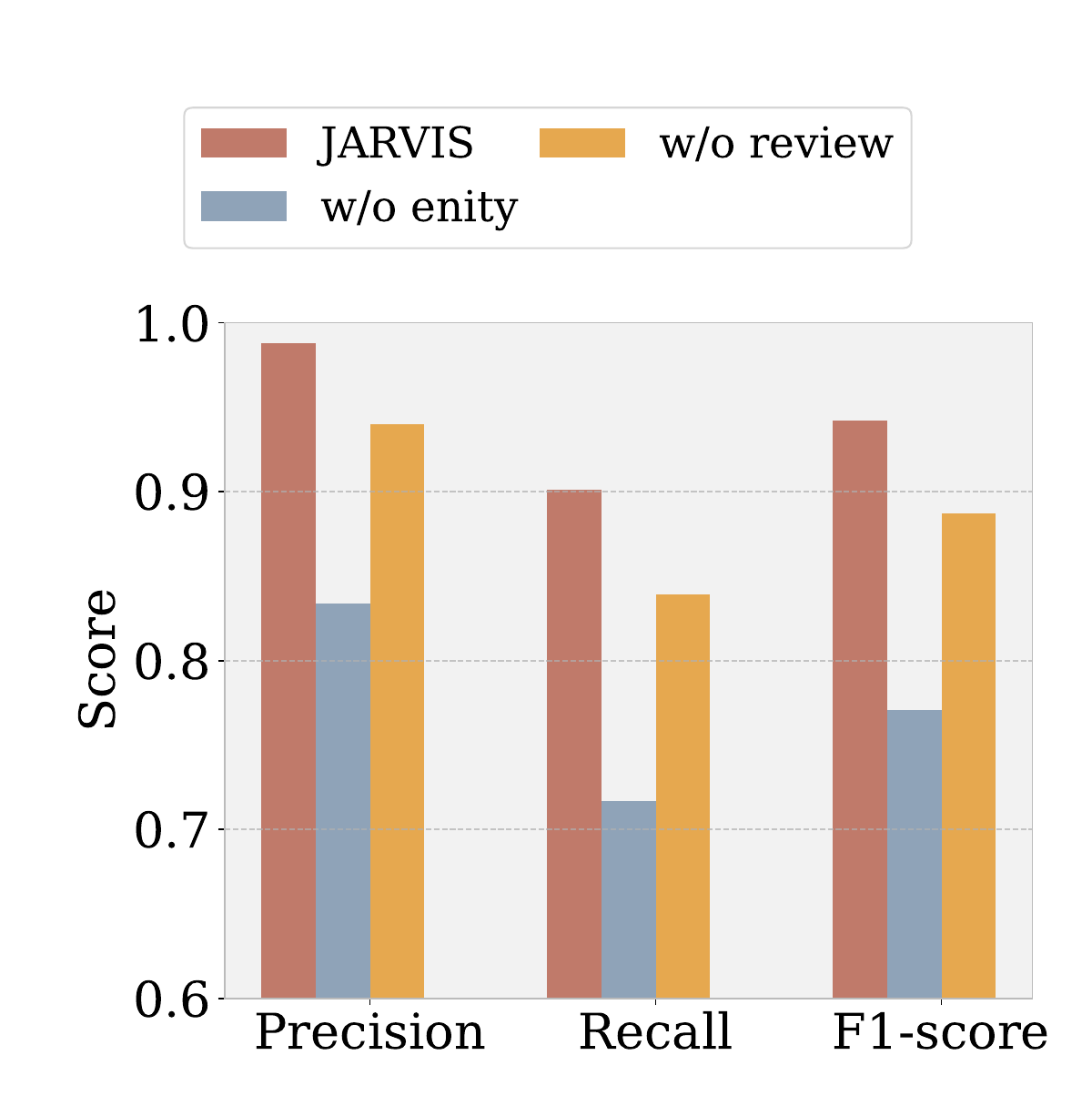}
    \caption{Graph Node Types.}
    \label{fig:ablation-b}
\end{subfigure}

\caption{Ablation Study on JARVIS components. "Dense" and "Sparse" denote \textit{Dense Embedding} and \textit{Sparse Embedding} relatively. "Review" and "Entity" denote \textit{Review Node} and \textit{Entity Node} in the evidence subgraph, respectively.}
\vspace{-1.0em}
\label{fig:ablation}
\end{figure}

We evaluate the model by individually removing review and entity nodes from the evidence subgraph (along with their corresponding edges). The results, as illustrated in Figure \ref{fig:ablation-b}, demonstrate that both node types and their associated relations are crucial for the reasoning process. Notably, the removal of entity nodes exerts a more substantial impact on evaluation metrics.
\vspace{-4pt}

\subsection{Online Deployment}
We deploy JARVIS to a production environment covering the entire JD.com platform where more than 1.5 million newly added reviews are processed into embeddings and integrated into the database every day. To optimize resource utilization, we introduce an inspection mechanism that triggers the pipeline only when anomalies are detected in review-related metrics or when manual verification is requested.

To provide a more comprehensive evaluation, we conduct an A/B test. The results demonstrate that JARVIS outperforms the existing baseline by \textbf{2.2\%} in manual verification precision and achieves a \textbf{27\%} increase in recall volume. Furthermore, the average time required for manual verification is reduced by \textbf{75\%}. Additionally, we evaluate the adoption rate of the LLM-generated reasoning. An analysis is deemed "adopted" if the verdict on the target review is accurate and the evidence chain derived from the heterogeneous subgraph provides a comprehensive and persuasive justification of the fraudulent behavior. Our evaluation reveals an exceptionally high adoption rate of \textbf{96.4\%}. All performance improvements are statistically significant with $p\text{-value} < 0.05$.
% \begin{table}[!htbp]
% \centering
% \footnotesize

% \setlength{\tabcolsep}{5pt}
% \renewcommand{\arraystretch}{1.1} 

% \resizebox{0.9\columnwidth}{!}{   
% \begin{tabular}{lcccc}
% \hline
% Method & Precision & MRT & Recall & EAR \\
% \hline
% Baseline & 0.956 & 6.2min & - & - \\
% JARVIS      & \textbf{0.978} & \textbf{1.4min} & \textbf{+27\%} & \textbf{0.964} \\
% \hline
% \end{tabular}
% }
% \caption{Performance Comparison Between Baseline and our proposed framework}
% \vspace{-1.6em}
% \label{tab:ab-test}
% \end{table}

\section{Conclusion}
In this work, we present JARVIS, an evidence-grounded adjudication system that integrates multimodal retrieval, heterogeneous evidence graph expansion, and LLM-based reasoning for deceptive review detection. By grounding risk judgments in retrieved semantic and relational evidence, JARVIS enables interpretable adjudication beyond traditional review-level or graph-level detection approaches. Offline experiments and real-world deployment demonstrate its effectiveness in uncovering residual risks and improving governance efficiency. We hope our methodology can serve as a robust blueprint for future research in constructing efficient and comprehensive fraud detection frameworks. 
Ultimately, we envision this work contributing to the cultivation of a more trustworthy and sustainable ecosystem for e-commerce platforms.

% \begin{acks}
% This work was supported by National Natural Science Foundation of China under Grant 62571026 and Beijing Natural Science Foundation under Grant L242082.
% \end{acks}

\section*{Presenter Bio}
Leyang Li is an algorithm engineer at JD.com, working on transaction risk control. His research focuses on fraud detection and generative AI. His email address is lileyang.tokki@jd.com.
\bibliographystyle{ACM-Reference-Format}
\balance
\bibliography{main}

@article{tufail2022effect,
  title={The effect of fake reviews on e-commerce during and after Covid-19 pandemic: SKL-based fake reviews detection},
  author={Tufail, Hina and Ashraf, M Usman and Alsubhi, Khalid and Aljahdali, Hani Moaiteq},
  journal={Ieee Access},
  volume={10},
  pages={25555--25564},
  year={2022},
  publisher={IEEE}
}

@article{alsubari2022data,
  title={Data analytics for the identification of fake reviews using supervised learning},
  author={Alsubari, S Nagi and Deshmukh, Sachin N and Alqarni, A Abdullah and Alsharif, Nizar and Aldhyani, TH and Alsaade, F Waselallah and Khalaf, Osamah I},
  journal={Computers, Materials \& Continua},
  volume={70},
  number={2},
  pages={3189--3204},
  year={2022}
}

@article{abdulqader2022fake,
  title={Fake online reviews: A unified detection model using deception theories},
  author={Abdulqader, Mujahed and Namoun, Abdallah and Alsaawy, Yazed},
  journal={IEEE Access},
  volume={10},
  pages={128622--128655},
  year={2022},
  publisher={IEEE}
}

@article{jing2022semi,
  title={Semi-supervised fake reviews detection based on AspamGAN},
  author={Jing-Yu, Chen and Ya-Jun, Wang},
  journal={Journal of Artificial Intelligence and Capsule Networks},
  volume={4},
  number={1},
  pages={17--36},
  year={2022}
}

@article{khurshid2018enactment,
  title={Enactment of ensemble learning for review spam detection on selected features},
  author={Khurshid, Faisal and Zhu, Yan and Xu, Zhuang and Ahmad, Mushtaq and Ahmad, Muqeet},
  journal={International Journal of Computational Intelligence Systems},
  volume={12},
  number={1},
  pages={387--394},
  year={2018},
  publisher={Springer}
}

@article{jalther2019reputation,
  title={Reputation reporting system using text based classification},
  author={Jalther, Divyanshu and Priya, G},
  journal={International Journal of Innovative Technology and Exploring Engineering},
  volume={8},
  number={8},
  pages={1555--1558},
  year={2019}
}

@article{khatun2025leveraging,
  title={Leveraging Big Data Frameworks for Spam Detection in Amazon Reviews},
  author={Khatun, Mst Eshita and Akter, Halima and Rehan, Tasnimul and Ahmed, Toufiq},
  journal={arXiv preprint arXiv:2509.21579},
  year={2025}
}

@article{csenol2025domain,
  title={Domain Knowledge-Enhanced LLMs for Fraud and Concept Drift Detection},
  author={{\c{S}}enol, Ali and Agrawal, Garima and Liu, Huan},
  journal={arXiv preprint arXiv:2506.21443},
  year={2025}
}

@article{liu2022detection,
  title={Detection of spam reviews through a hierarchical attention architecture with N-gram CNN and Bi-LSTM},
  author={Liu, Yuxin and Wang, Li and Shi, Tengfei and Li, Jinyan},
  journal={Information Systems},
  volume={103},
  pages={101865},
  year={2022},
  publisher={Elsevier}
}

@inproceedings{crawford2021using,
  title={Using inductive transfer learning to improve hotel review spam detection},
  author={Crawford, Michael and Khoshgoftaar, Taghi M},
  booktitle={2021 IEEE 22nd international conference on information reuse and integration for data science (IRI)},
  pages={248--254},
  year={2021},
  organization={IEEE}
}

@article{mohawesh2024fake,
  title={Fake review detection using transformer-based enhanced LSTM and RoBERTa},
  author={Mohawesh, Rami and Salameh, Haythem Bany and Jararweh, Yaser and Alkhalaileh, Mohannad and Maqsood, Sumbal},
  journal={International Journal of Cognitive Computing in Engineering},
  volume={5},
  pages={250--258},
  year={2024},
  publisher={Elsevier}
}

@article{gambetti2023combat,
  title={Combat ai with ai: Counteract machine-generated fake restaurant reviews on social media},
  author={Gambetti, Alessandro and Han, Qiwei},
  journal={arXiv preprint arXiv:2302.07731},
  year={2023}
}

@inproceedings{veluru2025multimodal,
  title={Multimodal Detection of Fake Reviews using BERT and ResNet-50},
  author={Veluru, Suhasnadh Reddy and Erukude, Sai Teja and Marella, Viswa Chaitanya},
  booktitle={2025 4th International Conference on Innovative Mechanisms for Industry Applications (ICIMIA)},
  pages={877--882},
  year={2025},
  organization={IEEE}
}

@article{liu2019opinion,
  title={Opinion spam detection by incorporating multimodal embedded representation into a probabilistic review graph},
  author={Liu, Yuanchao and Pang, Bo and Wang, Xiaolong},
  journal={Neurocomputing},
  volume={366},
  pages={276--283},
  year={2019},
  publisher={Elsevier}
}

@article{manaskasemsak2023fake,
  title={Fake review and reviewer detection through behavioral graph partitioning integrating deep neural network},
  author={Manaskasemsak, Bundit and Tantisuwankul, Jirateep and Rungsawang, Arnon},
  journal={Neural Computing and Applications},
  volume={35},
  number={2},
  pages={1169--1182},
  year={2023},
  publisher={Springer}
}

@inproceedings{ren2022research,
  title={Research on fake reviews detection based on graph neural network},
  author={Ren, Xunyi and Yuan, Ziyan and Huang, Jiaming},
  booktitle={International symposium on computer applications and information systems (ISCAIS 2022)},
  volume={12250},
  pages={290--297},
  year={2022},
  organization={SPIE}
}

@article{he2022detecting,
  title={Detecting fake-review buyers using network structure: Direct evidence from Amazon},
  author={He, Sherry and Hollenbeck, Brett and Overgoor, Gijs and Proserpio, Davide and Tosyali, Ali},
  journal={Proceedings of the National Academy of Sciences},
  volume={119},
  number={47},
  pages={e2211932119},
  year={2022},
  publisher={National Academy of Sciences}
}

@inproceedings{wang2025saft,
  title={SAFT: Structure-aware Transformers for Textual Interaction Classification},
  author={Wang, Hongtao and Yang, Renchi and Wang, Hewen and Zheng, Haoran and Xu, Jianliang},
  booktitle={Proceedings of the 48th International ACM SIGIR Conference on Research and Development in Information Retrieval},
  pages={771--781},
  year={2025}
}

@inproceedings{zhang2025dual,
  title={Dual-channel heterophilic message passing for graph fraud detection},
  author={Zhang, Wenxin and Zhong, Jingxing and Yao, Guangzhen and Han, Renda and Lin, Xiaojian and Jiang, Lei and Zhang, Zeyu and Luo, Cuicui},
  booktitle={2025 International Joint Conference on Neural Networks (IJCNN)},
  pages={1--8},
  year={2025},
  organization={IEEE}
}

@article{wang2025mmsrarec,
  title={MMSRARec: Summarization and Retrieval Augumented Sequential Recommendation Based on Multimodal Large Language Model},
  author={Wang, Haoyu and Wang, Yitong and Wang, Jining},
  journal={arXiv preprint arXiv:2512.20916},
  year={2025}
}

@inproceedings{liu2025multimodal,
  title={Multimodal semantic retrieval for product search},
  author={Liu, Dong and Ramos, Esther Lopez},
  booktitle={Companion Proceedings of the ACM on Web Conference 2025},
  pages={2170--2175},
  year={2025}
}

@inproceedings{liang2025embedding,
  title={Embedding-based Retrieval in Multi-Modal Content Moderation},
  author={Liang, Hanzhong and Shi, Jinghao and Shen, Xiang and Wang, Zixuan and Wen, Vera and Mehrani, Ardalan and Chen, Zhiqian and Wu, Yifan and Zhang, Zhixin},
  booktitle={Proceedings of the 48th International ACM SIGIR Conference on Research and Development in Information Retrieval},
  pages={4264--4268},
  year={2025}
}

@article{yang2022chinese,
  title={Chinese clip: Contrastive vision-language pretraining in chinese},
  author={Yang, An and Pan, Junshu and Lin, Junyang and Men, Rui and Zhang, Yichang and Zhou, Jingren and Zhou, Chang},
  journal={arXiv preprint arXiv:2211.01335},
  year={2022}
}

@article{chen2024bge,
  title={Bge m3-embedding: Multi-lingual, multi-functionality, multi-granularity text embeddings through self-knowledge distillation},
  author={Chen, Jianlv and Xiao, Shitao and Zhang, Peitian and Luo, Kun and Lian, Defu and Liu, Zheng},
  journal={arXiv preprint arXiv:2402.03216},
  volume={4},
  number={5},
  year={2024}
}

@article{yang2025qwen3,
  title={Qwen3 technical report},
  author={Yang, An and Li, Anfeng and Yang, Baosong and Zhang, Beichen and Hui, Binyuan and Zheng, Bo and Yu, Bowen and Gao, Chang and Huang, Chengen and Lv, Chenxu and others},
  journal={arXiv preprint arXiv:2505.09388},
  year={2025}
}

@article{tadelis2016reputation,
  title={Reputation and feedback systems in online platform markets},
  author={Tadelis, Steven},
  journal={Annual review of economics},
  volume={8},
  number={1},
  pages={321--340},
  year={2016},
  publisher={Annual Reviews}
}

@article{gossling2018manager,
  title={The manager's dilemma: a conceptualization of online review manipulation strategies},
  author={G{\"o}ssling, Stefan and Hall, C Michael and Andersson, Ann-Christin},
  journal={Current issues in Tourism},
  volume={21},
  number={5},
  pages={484--503},
  year={2018},
  publisher={Taylor \& Francis}
}

@article{lee2018sentiment,
  title={Sentiment manipulation in online platforms: An analysis of movie tweets},
  author={Lee, Shun-Yang and Qiu, Liangfei and Whinston, Andrew},
  journal={Production and Operations Management},
  volume={27},
  number={3},
  pages={393--416},
  year={2018},
  publisher={SAGE Publications Sage CA: Los Angeles, CA}
}

@article{Qwen3-VL,
      title={Qwen3-VL Technical Report}, 
      author={Shuai Bai and Yuxuan Cai and Ruizhe Chen and Keqin Chen and Xionghui Chen and others},
	  journal={arXiv preprint arXiv:2511.21631},
      year={2025}
}

@inproceedings{li2019spam,
  title={Spam review detection with graph convolutional networks},
  author={Li, Ao and Qin, Zhou and Liu, Runshi and Yang, Yiqun and Li, Dong},
  booktitle={Proceedings of the 28th ACM international conference on information and knowledge management},
  pages={2703--2711},
  year={2019}
}

@article{liu2025detecting,
  title={Detecting LLM-Generated Spam Reviews by Integrating Language Model Embeddings and Graph Neural Network},
  author={Liu, Xin and Xu, Rongwu and Jia, Xinyi and Liao, Jason and Sun, Jiao and Huang, Ling and Xu, Wei},
  journal={arXiv preprint arXiv:2510.01801},
  year={2025}
}


\end{document}